# Smart Waste Management System for Makkah City using Artificial Intelligence and Internet of Things

Rawabi Saeed Al Qurashi[1], Maram Meshal Almnjomi[1], Teef Lahg Alghamdi[1], Amjad Hasan Almalki[1], Shahad Saed Alharthi [1], Shahad Mashal althobut[1]i, Alanoud Saad Alharthi[1], Maha A. Thafar[1*]

[1] College of Computers and Information Technology, Computer Science Department, Taif University, Taif, Kingdom of Saudi Arabia.
corresponding author: `m.thafar@tu.edu.sa`, https://orcid.org/0000-0003-0539-7361

**Abstract.** Waste management is a critical global issue with significant environmental and public health implications. It has become more destructive during large-scale events such as the annual pilgrimage to Makkah, Saudi Arabia, one of the world's largest religious gatherings. This event's popularity has attracted millions worldwide, leading to significant and unpredictable accumulation of waste. Such a tremendous number of visitors leads to increased waste management issues at the Grand Mosque and other holy sites, highlighting the need for an effective solution other than traditional methods based on rigid collection schedules.

To address this challenge, this research proposed an innovative solution that is context-specific and tailored to the unique requirements of pilgrimage season: a Smart Waste Management System, called TUHR, that utilizes the Internet of Things and Artificial Intelligence. This system encompasses ultrasonic sensors that monitor waste levels in each container at the performance sites. Once the container reaches full capacity, the sensor communicates with the microcontroller, which alerts the relevant authorities. Moreover, our system can detect harmful substances such as gas from the gas detector sensor. Such a proactive and dynamic approach promises to mitigate the environmental and health risks associated with waste accumulation and enhance the cleanliness of these sites. It also delivers economic benefits by reducing unnecessary gasoline consumption and optimizing waste management resources. Importantly, this research aligns with the principles of smart cities and exemplifies the innovative, sustainable, and health-conscious approach that Saudi Arabia is implementing as part of its Vision 2030 initiative.

## 1   Introduction

Waste management is a critical function of urban sustainability, with far-reaching implications for public health, the environment, and economic efficiency [1]–[3]. It encompasses the systematic collection, transportation, and waste processing to maintain clean and healthy habitable areas [4]. Waste management becomes notably complex in large-scale events such as the annual Hajj pilgrimage to Makkah, one of the most populated cities in Saudi Arabia. This pilgrimage attracts millions of people each year, leading to substantial waste production. Traditional systems reliant on preset collection schedules, are inefficient in managing the extremely high volume and unpredictability of the waste generated at the different holy sites. The collection schedules may either be too early, resulting in inefficient pickups from yet-to-be-filled bins, or too late, leading to bin overflows [5].

In recent years, the rapid development of technologies such as internet of things (IoT) and artificial intelligence (AI) including machine learning (ML) and deep learning (DL), has opened up new opportunities in different sectors, including healthcare and medicine [6]–[9], transportation [10], robotics [11], [12], energy [13], and smart cities [14]. These innovations hold the potential to enhance the effectiveness of systems across these domains. In the context of waste management, these technologies promise to enhance the efficiency of waste collection and transportation services, which are vital components of municipal cleaning programs [15], [16]. IoT-enabled smart sensors and communication systems offer the potential to detect real-time bin fill levels and schedule collections, accordingly, preventing premature pickups or delays. Several recent approaches have suggested implementing technologies like IoT to improve waste management through smart sensors and communication systems [17].

This research focuses on the design and implementation of the innovative waste detection and management system at the holy city of Makkah during the Hajj season. We aim to enhance the cleanliness, healthiness, and general experience of one of the largest religious gatherings in the world by applying state-of-the-art technologies. This work aligns with the strategic objectives of Saudi Arabia's Vision 2030 [18]. The focus of this vision is to have an environmentally sustainable guide, a vibrant society, and an innovative economy. Although our immediate focus is to improve the Hajj experience, our greater aim is to contribute to realizing Saudi Arabia's Vision 2030. This study contributes to waste management field in several significant ways, as follows:

1. **Focus on Makkah's Unique Context:** by addressing the challenges faced in Makkah during the Hajj season. The proposed system is designed to handle the unpredictable waste generated during this time, providing a solution tailored to Makkah's unique needs. This aligns with Vision 2030's goal of improving services for the Hajj and Umrah pilgrimages.
2. **Accurate and Efficient Waste Collection:** the system will ensure time-efficient waste collection, prevent bin overfills, and improve the city's cleanliness, aligning with the vision's objective of enhancing the urban environment.
3. **Healthier environment:** By preventing bin overfills and optimizing waste management, the study contributes to a healthier environment in Makkah, particularly during crowded seasons.
4. **Detection of harmful substances:** Using a gas detector sensor gives our system the extra ability to monitor the existence of harmful substances, such as gas leaks from the fire. Thus, workers can know this and intervene quickly in emptying the containers and extinguishing these gases.

The overall structure of this paper is as follows. The first section is the introduction, which provides a problem overview, motivation, and the contribution of this research. Section 2 discusses the literature review. Section 3 presents the methodology with all the details. Section 4 revolves around system implementation and the experiment's simulation and results. Finally, we conclude this work, highlight some limitations, and recap several potential future directions in section 5.

## 2      Background and Related Works

Recently, there has been a growing focus on research aimed at developing smart waste management systems. Several studies have been published to address this issue by leveraging AI and IoT technologies [16], [17], [19], [20]. These works highlight integrating smart technologies and IoT solutions to enhance waste management efficiency and sustainability in Makkah [20]–[22] and other cities worldwide [1], [23]–[25].

One of the studies was done by Kamm and coauthors [26], where they designed a case study to investigate the efficacy of smart waste management systems. They installed sensors in glass bins to measure the fill level of each bin. Their system included an ultrasonic sensor with a real-time clock and a battery capable of scheduling deep-sleep modes. They also used a LoRa chip for long-range data transmission. The software comprised a sensor application, the things network service, and a web application. The study presented the following constraints: city topology, network coverage and penetration, and energy consumption for data transmission. Another study [27] discussed waste collection in smart cities based on IoT. The implementation included hardware such as HC-SR04, an ultrasonic sensor to measure bin fill levels, and a GPS module to track each bin's location. The data from the sensors and GPS modules were communicated via a Raspberry Pi microcontroller to a cloud platform for storage and analysis. The cloud would then signal an Android application. An interesting feature is that after emptying each waste bin, a water pump was activated to clean the bin.

A different study [28] proposed a smart waste management system and demonstrated its effectiveness through a real-life case study. The IoT module's middleware, which included a decision unit, database, and management unit, served as the connecting hub for the "My Waste Bin" module (which included sensors like the HC-SR04 and devices such as the DHT11 temperature and humidity sensor and GPS), and the "My Waste App" module. The system was powered by an external rechargeable battery attached to a photovoltaic solar panel. However, the system lacked the ability to calculate the most efficient route for waste collection trucks. Another recent study published in 2022 in [22] focused on organic waste management in Saudi Arabia. The study proposed an integrated system including educating the community, public participation, policy development, and advanced technology. The system utilized IoT cloud, ultrasonic sensors, GPS, Arduino, GPRS, and Google Maps. The waste management process began with identifying waste sources, followed by smart bin-enhanced waste collection, transportation, and disposal. The smart waste collection is enabled through smart bins that send data to the management center (MC) via GSM or GPRS. The MC dispatched waste collection trucks when the bin signals its fullness. In addition, the suitable collection truck is chosen based on the weight that it can handle, and a suitable route is calculated through a smart routing algorithm, namely the Genetic Algorithm. This becomes possible because the bin ID and location are communicated with the MC. The study also emphasized the potential of organic waste as a renewable energy source and advocated for community education and supportive laws and policies.

The last study was conducted by Abdullah and coauthors [20], who proposed a smart waste management solution for Makkah to address the rapid accumulation of waste contributing to air and water pollution. The system was designed to operate at three levels: the main city, the municipalities, and the district management centers. Each smart bin and truck were equipped with ultrasonic sensors, Arduino, GPRS, and GPS, all connected through a cloud server. The IoT-enabled system included a monitoring application that allowed the smart bins to report their status and request collection once they reached 90% capacity, while also identifying the type of waste they contained. Furthermore, the smart truck was notified based on the received location, available free space, and waste type. Additionally, the system featured a manual waste collection option, where the truck driver collected the waste and updated the data through the application.

Although significant research has been done on smart waste management systems, their full integration into operational settings, particularly in Makkah, is still under development. Many existing systems lack scalability, real-time responsiveness, and large-scale integration with city management. Therefore, there is a pressing need to transition from theoretical solutions to practical deployment. Collaboration with local authorities is crucial for adapting these technologies to Makkah's specific challenges, ensuring efficient and sustainable waste collection, and improving both environmental and public health outcomes.

## 3 Methodology

The methodology presents the feasibility study. Then it outlines the approaches that will be employed to design and implement the smart waste management system. The methodology went through a series of stages, including understanding the system architecture, applying IoT and AI techniques, and incorporating an administrative portal for system management. The objective is to ensure that the system is resilient, efficient, and user-friendly, capable of effectively managing waste problems during periods of high demand.

### 3.1 Feasibility Study

The feasibility study highlights the environmental and sociological importance of the TUHR Smart Waste Management System based on survey data. The survey included over 300 participants, with 57.2% from the population and 42.8% from visitors. Among the respondents, 95.2% agreed that improved waste management during the Hajj season is crucial, and 95.5% supported the implementation of a smart waste management system in Makkah. Survey results also revealed that 87.8% of respondents noticed frequent waste accumulation during the Hajj season, emphasizing the need for timely waste collection. Additionally, 93.2% believed that smart containers could reduce pollution from waste transport truck exhausts. Moreover, 41% of respondents indicated that waste collection should occur twice a day during the Hajj period, with 38.5% supporting three times a day collection.

Among officials at the Grand Mosque and holy sites, 98.7% identified crowding as a major waste management challenge, and 94.9% pointed out the insufficiency of bins and ineffective waste collection methods. These findings support the need for a smart waste management system that can address these challenges.

### 3.2 Problem Formulation

In this study, the problem can be systematically formulated as follows:
- **Input:** The input involves the status information for each waste bin, indicating whether it is full or empty, which is collected through IoT-enabled smart bins equipped with sensors. Also, the geographic location of each bin within the city of Makkah is necessary, which can be obtained using a GPS technology.
- **Output:** The output is a time-specific signal indicating when waste collection should occur and an optimized route for sanitation workers to follow.
- **Constraints:** The constraints include ensuring that each full bin is assigned to only one worker to avoid redundancy. Furthermore, the distance traveled by each worker should be minimized to increase efficiency and reduce fuel consumption. This is achieved through the intelligent route optimization process powered by AI and Google map.
- **Goal**: to ensure timely and effective waste collection and to optimize sanitation workers' routes.

### 3.3 System Overview

This section describes our proposed Smart Waste Detection and Management System's architecture, called TUHR. Fig. 1 illustrates the TUHR workflow. The Flutter mobile application serves as the interface between the users (i.e., workers) and the system server.

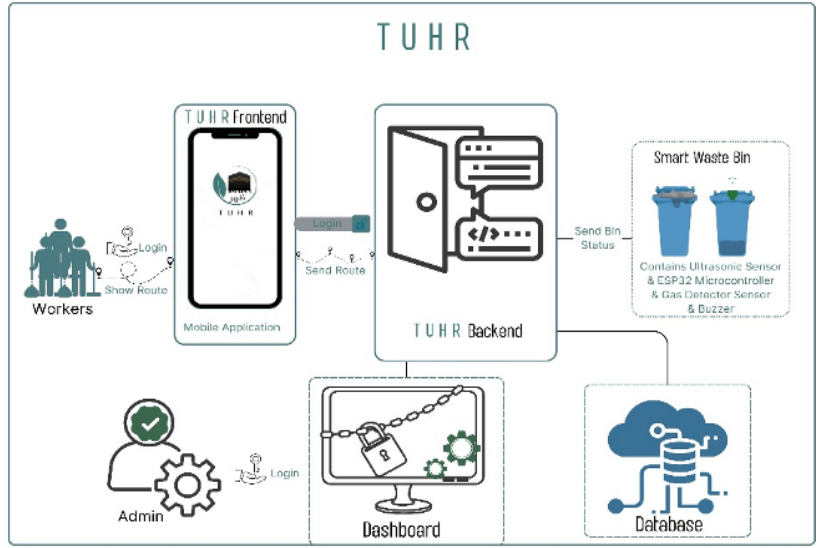

**Fig. 1**: Workflow of the proposed smart waste system.

The TUHR System workflow can be summarized in the following steps:

1. **Data Collection:** Each smart bin is equipped with an ESP32 microcontroller and an ultrasonic sensor. The sensor monitors the bin's fill level continuously, and the microcontroller transmits this data to a cloud server in real time.
2. **Data Processing and Storage:** The cloud servers will receive the data from the smart bins, process it, and store it in the database.
3. **Full Bin Detection and Alerting:** The system continually checks the fill status of each bin. When a bin reaches its capacity, the system triggers an alert to notify the relevant users and initiates the process of planning waste collection and calculates the most efficient route.
4. **Route Optimization:** The AI algorithm determines the optimal path for the worker to reach the designated bin using Google Map application.
5. **Notification Dispatch:** workers receive notifications about the bin status and its location.
6. **Bin Status Update:** Once the waste has been collected, the bin's status in the system is updated to "empty." This change in status is reflected in real time across the system, ensuring all users have access to current information.

## 4　Experiments and Results

This section discusses the simulation of the TUHR system, explaining the hardware setup and configuration, discusses implementing of TUHR application, building the admin dashboard, and integrating those system components.

### 4.1 TUHR Software Implementation

The TUHR system consists of two main software components: the first one is **TUHR application for workers.** The application is designed to cater to the needs of "Worker" employees by allowing them to modify their personal information and efficiently perform the required tasks. Workers can log into the application using their username and password, which were previously registered in the system database. The second component is **administrative dashboard,** which allows system administrators to manage the entire system through a centralized web interface. The administrator gains access to the dashboard by entering the designated username and password in the administrator login interface. Once logged in, the initial page that appears is the "Dashboard," which contains several sections (see **Figure 2**). These sections include "My Profile," "Manage Users," "Manage Zones," "Manage Sensors," "View Reads," and "Contacts." The administrator has full access to these pages and can navigate them as needed. After completing their tasks, the administrator can log.

### 4.2 Integration of Hardware and Software Components

Initially, the hardware and software were designed separately based on the specific requirements of the smart waste management system, TUHR. Next, these components were integrated to ensure the smooth and accurate function of the system. This integration process involves several key factors: physically connecting the hardware components within the smart waste containers, enabling efficient and continuous communication, synchronizing timing, and establishing a unified database. The database serves as a central hub, ensuring that both the dashboard and the mobile application receive the same data from the sensors simultaneously. The hardware and software components are integrated to ensure seamless communication and compatibility across devices. Figure 2 illustrates the integration between the system components and how the dashboard and mobile application function simultaneously.

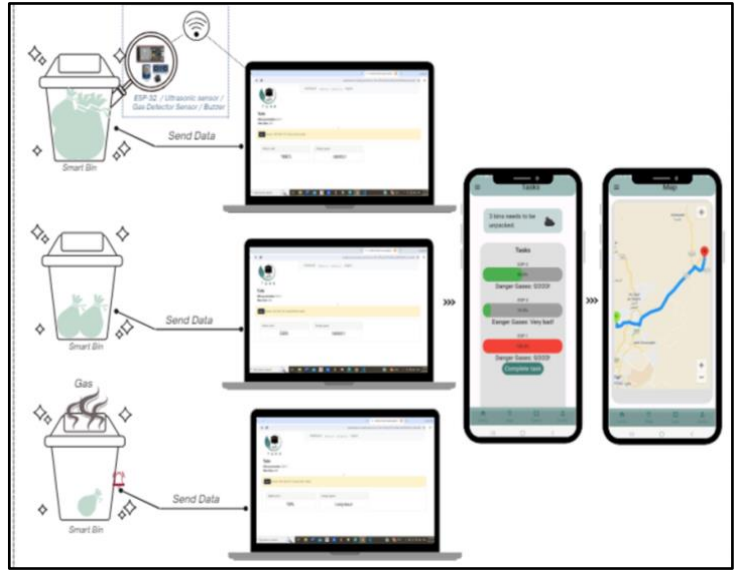

**Fig. 2**: Devices Integration with Dashboard and TUHR Application.

### 4.3 Experiment Setup and Procedure

This section covers a series of small experiments that were carried out to benchmark the system's performance related to waste management. The system's setup is described, with a discussion of the results. The aim of the experiments is to show the validity of the system in realistic situations and to test the system usability.

**Initial Step and Configuration**

Figure 3 depicts the integration of three key components: the ultrasonic sensor, the gas detector, and the buzzer, with the microcontroller on the electrical circuit board. Each component is programmed with specific code to perform its designated task. After the code was uploaded, it was thoroughly tested to ensure that each component functions correctly. The initial setup involved three medium-sized containers; each container was equipped with an electrical with the three components.

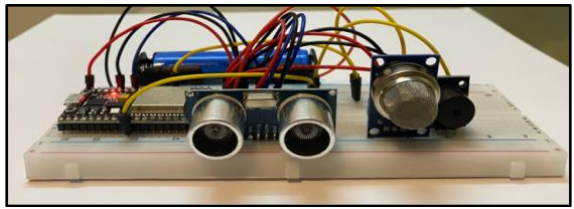

**Fig. 3**: Electrical Circuit

**Experiments Scenarios and Evaluation**

We conducted a series of small-scale experiments to perform real-world operations with the TUHR System. **First Experiment Scenario:** To simulate different waste levels (e.g., 0% "empty," 50% "almost full," and 95% "full"), each container was filled with varying amounts of waste (see Figure 4). This setup allowed us to demonstrate the system's capability to accurately estimate and display waste levels on the application's dashboard.

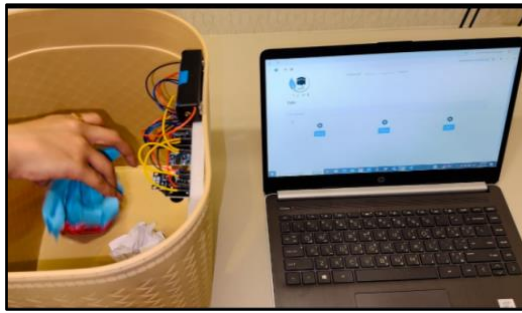

**Fig. 4:** First Experiment Scenario with administration dashboard showing in the screen.

**The Second Experiment Scenario** (Figure 5): we conducted a fire experiment to test the functionality of the gas sensor and the buzzer alarm, ensuring that both components activated appropriately in response to simulated hazardous conditions.

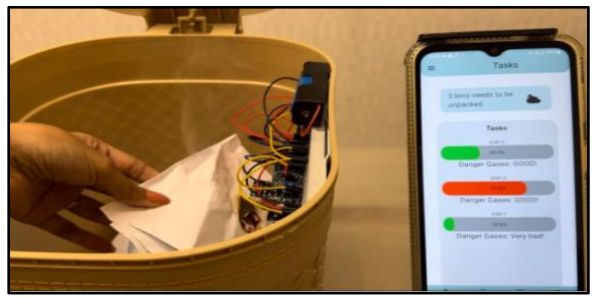

**Fig. 5:** Second Experiment Scenario of detecting fire, with mobile application notification.

## 5   Conclusion

The TUHR Smart Waste Management System has shown significant promise in enhancing waste management efficiency during high-demand events like the Hajj pilgrimage. The system's use of real-time data transmission and smart sensors allows for better coordination of waste collection activities, which addresses common issues faced by traditional systems, such as premature or delayed pickups. Previous research on smart waste management, also emphasizes the role of IoT in improving waste management efficiency, especially in large-scale urban settings. However, unlike many existing systems that focus on specific applications or geographic areas, TUHR is tailored to the unique conditions of Makkah during the Hajj, a context that has not been widely studied in the existing literature.

One key strength of the TUHR system is its potential to prevent bin overflows, which have been a persistent issue in large-scale events. By leveraging IoT sensors, the system can accurately monitor fill levels and schedule collections in real time. This approach has the potential to significantly reduce inefficiencies, which are common in traditional systems, as highlighted in earlier studies. The integration of AI ML models further contributes to the system's effectiveness by allowing for data-driven decision-making that improves with time and usage.

A key limitation was the difficulty in obtaining real-world data for accurate testing, which impacted the system's evaluation under actual conditions and not fully reflect real-world complexities. The integration of multiple hardware components and ensuring smooth communication among them, especially in dynamic, large-scale environments, posed significant technical challenges. Additionally, the system's reliance on network connectivity, which can fluctuate in crowded settings, limited its overall reliability. Despite these limitations, our system demonstrates potential for improving waste management practices. This involve future work such as managing the hardware integration, refining sensor calibration, and testing within the real operation scenario. Further enhancements in data processing and analytics performance to address large datasets lead to improved and scalable performances. This advancement will go the extra mile toward cleanliness and optimize waste management practices at holy sites.